\documentclass[superscriptaddress,
  aps,
  pra,
  amsmath,amssymb,
  reprint,
]{revtex4-1}
\usepackage{graphicx,color}
\usepackage{amsmath}
\usepackage{natbib}
\usepackage{epsfig}
\usepackage{graphicx}

\begin{document}

\title{\color{blue} Ion drift instability in a strongly coupled collisional complex plasma}

\author{Sergey Khrapak}
\email{Sergey.Khrapak@dlr.de}
\affiliation{Institut f\"ur Materialphysik im Weltraum, Deutsches Zentrum f\"ur Luft- und Raumfahrt (DLR), 82234 We{\ss}ling, Germany}
\affiliation{Joint Institute for High Temperatures, Russian Academy of Sciences, 125412 Moscow, Russia}

\author{Victoria Yaroshenko}
\affiliation{Institut f\"ur Materialphysik im Weltraum, Deutsches Zentrum f\"ur Luft- und Raumfahrt (DLR), 82234 We{\ss}ling, Germany}

\begin{abstract}
We investigate the low-frequency wave mode associated with heavy particles   and its instability in a collisional complex plasma with drifting ions. The effect of the ion drift on the sound velocity of this mode is discussed. The general condition of the instability is derived for subthermal ion drifts, taking into account strong coupling of the particle component. As a general tendency, strong coupling effects reduce the sound velocity and facilitate the occurrence of the ion drift instability. A wide parameter range is considered from the weakly collisional to strongly collisional regimes for the ion and particle components. The chosen plasma parameters are representative to the PK-4 experiment, currently operational on board the International Space Station.   
        
\end{abstract}

\date{\today}

\maketitle

\section{Introduction}

Complex (dusty) plasmas are multicomponent plasma systems, where one of the charged components represents a collection of massive and highly charged (dust) particles~\cite{TsytovichUFN1997,Bonitz2010,FortovUFN2004,FortovPR2005,
ShuklaRMP2009,ThomasPPCF2019}. 
The presence of these massive charged particles in a plasma introduces new low-frequency collective modes associated with the dynamics of the particle component. The longitudinal collective mode is usually referred to as the dust-acoustic wave (DAW)~\cite{ShuklaRMP2009,RaoDAW,MerlinoPoP1998,MerlinoJPP2014} (because of the acoustic-like dispersion at long wavelengths) and plays an extremely important role in complex plasma research. The DAW dispersion relation was originally derived for an isotropic complex plasma system, where both the electrons and ions follow the Boltzmann distribution in the wave potential, while particles react dynamically~\cite{RaoDAW}. In more complicated situations this low frequency mode is sometimes referred to more generically as the dust density wave (DDW)~\cite{Piel2011,MerlinoJPP2014}.    

Dust acoustic or density waves can be excited naturally due to various instabilities or produced by means of controlled action using specially designed devices (most common are electrical and laser radiation manipulations)~\cite{FortovPR2005}. The ion-dust streaming instability represents one of the main mechanisms of spontaneous excitations of low frequency paticle waves~\cite{MerlinoPoP2009}. In typical laboratory conditions, this instability originates from the ion drift through the particle cloud, produced by external electric fields that sustain the discharge. The ion-dust streaming instability conditions have been extensively analysed, first for the ideal weakly coupled complex plasma regime~\cite{RosenbergPSS1993,MelandsoJGR1993,RosenbergJVST1996,AngeloPSS1996,
JoycePRL2002} and, more recently, for the strongly coupled complex plasma regime~\cite{Rosenberg1998,KalmanJPA2003,RosenbergPRE2014}. It has been demonstrated that strong coupling, in general, leads to an enhancement of the instability growth rate~\cite{RosenbergPRE2014}. 

The purpose of this paper is to present a practical (and rather general) condition (threshold) of the ion-dust streaming instability in typical laboratory gas discharges. Our main motivation is relevant to the PK-4 complex plasma laboratory, currently operational onboard the International Space Station (ISS)~\cite{PK4}, where investigations of wave processes represent a significant part of the overall research program~\cite{JaiswalPoP2018,YaroshenkoPoP2019}. Taking into account specifics of these experiments we extend the analysis of Ref.~\cite{RosenbergPRE2014} to the regime where the ion mean free path with respect to collisions with neutrals is shorter than the observed wavelengths. Regarding the particle component, we take into account strong coupling effects and particle-neutral collisions. We focus on linear waves and sufficiently long wavelengths here, where the acoustic-like dispersion takes place. Various complications, such as particle charge variations, forces affecting particles (ion and electron drag, polarization force, etc.), ion wakes behind the particles, particle drifts, etc. are omitted to make the main problem tractable.  
 
\section{Plasma parameter regime}\label{parameters}

First, let us introduce the main dimensionless parameters which determine the dispersion relation and the instability threshold in the considered situation. The ion {\it thermal} Mach number is 
\begin{displaymath}
M=u/v_{{\rm T}i},
\end{displaymath}   
where $u$ is the ion drift velocity and $v_{{\rm T}i}=\sqrt{T_i/m_i}$ is the ion thermal velocity, with $T_i$ and $m_i$ being the ion temperature and mass, respectively. The ion collisionality parameter is
\begin{displaymath}
\theta_i=\nu_i/\omega_{{\rm p}i}\simeq \lambda_i/\ell_i,
\end{displaymath}
where $\nu_i$ is the ion-neutral collision frequency, $\omega_{{\rm p}i}=\sqrt{4\pi e^2 n_i/m_i}$ is the ion plasma frequency, $\lambda_i=v_{{\rm T}i}/\omega_{{\rm p}i}$ is the ion Debye radius, and $\ell_i$ is the ion mean free path with respect to collisions with neutrals. The standard notation is used: $e$ is the elementary charge, $n_i$ is the ion density, and ions are singly charged.

Similarly, the particle collisionality parameter is
\begin{displaymath}
\theta_p=\nu_p/\omega_{{\rm p}},
\end{displaymath}
where $\nu_p$ is the characteristic frequency of particle momentum loss in particle-neutral collisions, $\omega_{{\rm p}}=\sqrt{4\pi Q^2n_p/m_p}$ is the plasma frequency associated with the particle component, $Q$ is the particle charge, $n_p$ is the particle density, and $m_p$ is the particle mass. 
  
To get an idea about the parameter regime of interest here, consider the PK-4 laboratory on the ISS. Here the particle are immersed in a long discharge tube. The particle cloud can slowly drift or be trapped inside the tube, while the ions are drifting relative to the cloud due to the presence of an axial electric field, $E\sim{\mathcal O}$(V/cm), in the positive column of the discharge (see Ref.~\cite{PK4} for further details). In order to characterize the discharge, Langmuir probe measurements have been performed in the particle-free plasma under different experimental conditions in the laboratory~\cite{FortovPK4}. For a representative case of the discharge in neon at a current of 1 mA, the measured electron density, electron temperature, and electric field in the pressure range of $20-100$ Pa are plotted in Fig.~3 of Ref.~\cite{AntonovaPoP2019}. The pressure dependence can be relatively well described by simple linear fits~\cite{KhrapakPRE2005,KhrapakPRE2013}: $n_e\simeq n_i\simeq (0.9+0.03p_{[Pa]})\times 10^8$ (cm$^{-3}$), $T_e\simeq 8.3-0.02p_{[Pa]}$ (eV), and $E\simeq 2.1$ (V/cm), where $p_{[Pa]}$ is the pressure expressed in Pascals.

\begin{figure}
\centering
    \includegraphics[width=7 cm]{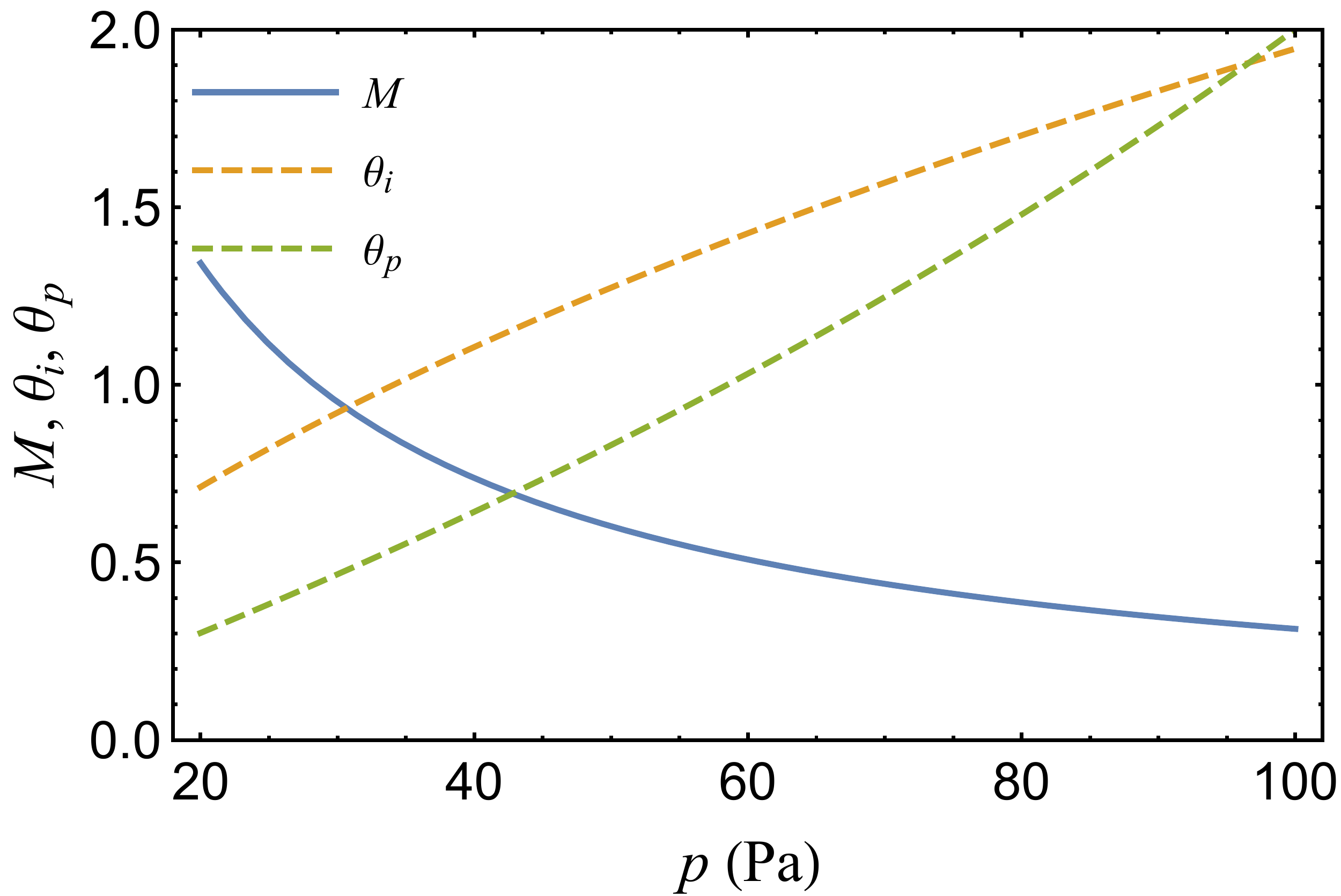} 
    \caption{Ion thermal Mach number $M$ along with the collisionality parameters $\theta_i=\nu_i/\omega_{{\rm p}i}$ and  $\theta_p=\nu_p/\omega_{{\rm p}}$ for the ion and particles components, respectively, versus the neutral gas pressure $p$. Calculation is for the conditions corresponding to the PK-4 discharge in neon with the electrical current of 1 mA.}\label{Fig1}
\end{figure}

The results of calculating $M$, $\theta_i$, and $\theta_p$ as functions of pressure are shown in Fig.~\ref{Fig1}. The thermal Mach number is calculated from a modified Frost formula~\cite{KhrapakFrost}. The effective ion-neutral collision frequency, which accounts for the ion drift effect, is then calculated from a simple practical expression proposed in Ref.~\cite{KhrapakJPP2013}. This, together with pressure dependence of $n_i$ defines the ion collisional ratio $\theta_i$. The particle-neutral collision frequency is evaluated using the standard Epstein drag formula~\cite{Epstein,LiuPoP2003}. The dust plasma frequency is estimated for a particle of $a=1.7$ $\mu$m radius, using a typical (constant) particle density $n_p\simeq 3\times 10^4$ cm$^{-3}$, and assuming a constant reduced charge of $z=|Q|e/aT_e\simeq 0.3$ (see Fig.~5 from Ref.~\cite{AntonovaPoP2019}). This allows us to evaluate the particle collisional ratio $\theta_p$.

The results are plotted in Fig.~\ref{Fig1}. We observe that the ion drift becomes subthermal for $p\gtrsim 30$ Pa. The ions become highly collisional ($\theta_i>1$) at approximately the same pressure. The Epstein frequency  remains remains smaller than the plasma-particle frequency ($\theta_p<1$) for $p\lesssim 60$ Pa. Concerning the characteristic length scales, the following hierarchy emerges. The screening length, which can be approximated by the ion Debye radius, varies in the range from $\lambda\simeq 100$ $\mu$m to $\lambda\simeq 70$ $\mu$m. The interparticle separation is usually a factor of a few longer. In the considered example with $n_p=3\times 10^4$ cm$^{-3}$, we get $\Delta=n_p^{-1/3}\simeq 300$ $\mu$m. The waves typically observed experimentally, have wavelengths of the order of a few millimetres, much longer than the interparticle separation. Consequently we have $k\lambda\lesssim k\Delta\ll 1$. This overall situation is representative for many complex plasma experiments in the laboratory (not only for the PK-4 laboratory) and will be investigated in detail in the remaining of this paper.

\section{Dispersion relation}

The plasma dielectric permittivity is
\begin{equation}\label{permit}
\epsilon(k,\omega)=1+\sum_j \chi_j\simeq \sum_j \chi_j,
\end{equation}
where the summation is over  charged species $j$. The last approximation applies to the long-wavelength limit, where quasineutrality condition holds (and the Laplacian in the Poisson equation can be neglected). We consider a weakly ionized complex plasma consisting of three charged component (electrons, ions, and charged particles) and one neutral component (neutral gas). For light mobile electrons it is sufficient to keep the static contribution
\begin{equation}\label{electronresp}
\chi_e=\frac{1}{k^2\lambda_e^2},
\end{equation} 
where $\lambda_e=\sqrt{T_e/4\pi e^2 n_e}$ is the electron Debye radius.  

The ion response is~\cite{AlexandrovBook} 
\begin{equation}\label{ionresp1}
\chi_i=\frac{1}{k^2\lambda_i^2}\frac{1+\xi_i Z(\xi_i)}{1+\frac{i\nu_i}{\sqrt{2}kv_{{\rm T}i}}Z(\xi_i)},
\end{equation}
where $\xi_i=\frac{\omega-{\bf k\cdot u}+i\nu_i}{\sqrt{2}kv_{{\rm T}i}}$, and $Z(x)$ is the plasma dispersion function~\cite{LL_KInetics},
\begin{displaymath}
Z(x)=\frac{1}{\sqrt{\pi}}\int_{-\infty}^{+\infty}\frac{e^{-\zeta^2}}{\zeta-x}d\zeta.
\end{displaymath}
For the considered low-frequency mode associated with the dynamics of the particle component we can safely neglect $\omega$ in the argument $\xi_i$, since both inequalities $\omega\ll |{\bf k\cdot u}|$ and $\omega\ll \nu_i$ are very well satisfied~\cite{JoycePRL2002}. Since we will consider propagation along the ion flow, we will simply use $ku$ instead of ${\bf k \cdot u}$.  

The generalized particle response is
\begin{equation}\label{partresp}
\chi_{\rm p}=-\frac{\omega_{\rm p}^2}{\omega(\omega+i\nu_p)-\omega_{\rm p}^2{\mathcal D}_{L}(k)},
\end{equation}
where 
\begin{equation}\label{QLCA}
\omega_{\rm p}^2{\mathcal D}_L(k)=\frac{n}{m}\int\frac{\partial^2 \phi(r)}{\partial z^2}[g(r)-1]\left[1-\cos(\bf{k\cdot z})\right]d{\bf r}
\end{equation}
is the contribution from particle-particle correlations within the quasilocalized-charge approximation (QLCA)~\cite{GoldenPoP2000,KalmanPRL2000,DonkoJPCM2008,KhrapakPoP2016,
KhrapakIEEE2018}. It is essentially the longitudinal projection of a dynamical matrix~\cite{RosenbergPRE2014} and is expressed via the pairwise interparticle interaction potential $\phi(r)$ and the radial distribution function (RDF) $g(r)$. The contribution from particle-particle correlations vanishes in the disordered gaseous regime with $g(r)=1$. The interparticle interaction potential is usually assumed to be of Yukawa (screened Coulomb) shape, but deviations from this shape can also be treated within the QLCA~\cite{Fingerprints}.  

The dispersion relation is determined from the condition $\epsilon(k,\omega)=0$. Let us now consider some simple limiting regimes and derive useful approximations, before deriving the general instability condition. First, consider the simplest situation, when ion-neutral and particle-neutral collisions, ion drift, and strongly coupled effects are all absent. Then we get, in the long-wavelength limit, the acoustic dispersion of the form
\begin{equation}\label{DAW}
\omega=\omega_{\rm p}\lambda k = c_{\rm DA}k,
\end{equation}     
where $c_{\rm DA}=\omega_{\rm p}\lambda$ is the conventional dust acoustic velocity~\cite{RaoDAW}, and $\lambda$ is the linearised Debye radius, $\lambda^{-2}=\lambda_e^{-2}+\lambda_i^{-2}$ (in the usual situation with $T_e\gg T_i$ we have $\lambda\simeq \lambda_i$, as mentioned earlier). 

If we keep the plasma collisionless and isotropic (no ion drift), but retain the strongly coupled effects, the real part of the dispersion relation becomes
\begin{equation}\label{sound}
\omega^2=\omega_{\rm p}^2\lambda^2 k^2+\omega_{\rm p}^2{\mathcal D}_L(k)=c_{\rm s}^2k^2,
\end{equation}
where $c_{\rm s}$ is the resulting sound velocity.
The strong coupling correction ${\mathcal D}_L(k)$ is negative and hence the sound velocity is reduced compared to the weakly coupled DA value, $c_{\rm s}<c_{\rm DA}$~\cite{KalmanPRL2000,DonkoJPCM2008,KhrapakPoP2016}.
This effect is small near the one-component plasma limit, when $\kappa=\Delta/\lambda<1$, but becomes very well pronounced as $\kappa$ increases~\cite{KhrapakPRE2015_Sound,KhrapakPPCF2016,KhrapakPoP2019}. A practical approach to estimate ${\mathcal D}_L(k)$ will be discussed in Section~\ref{D_L}.

\begin{figure}
\centering
    \includegraphics[width=7 cm]{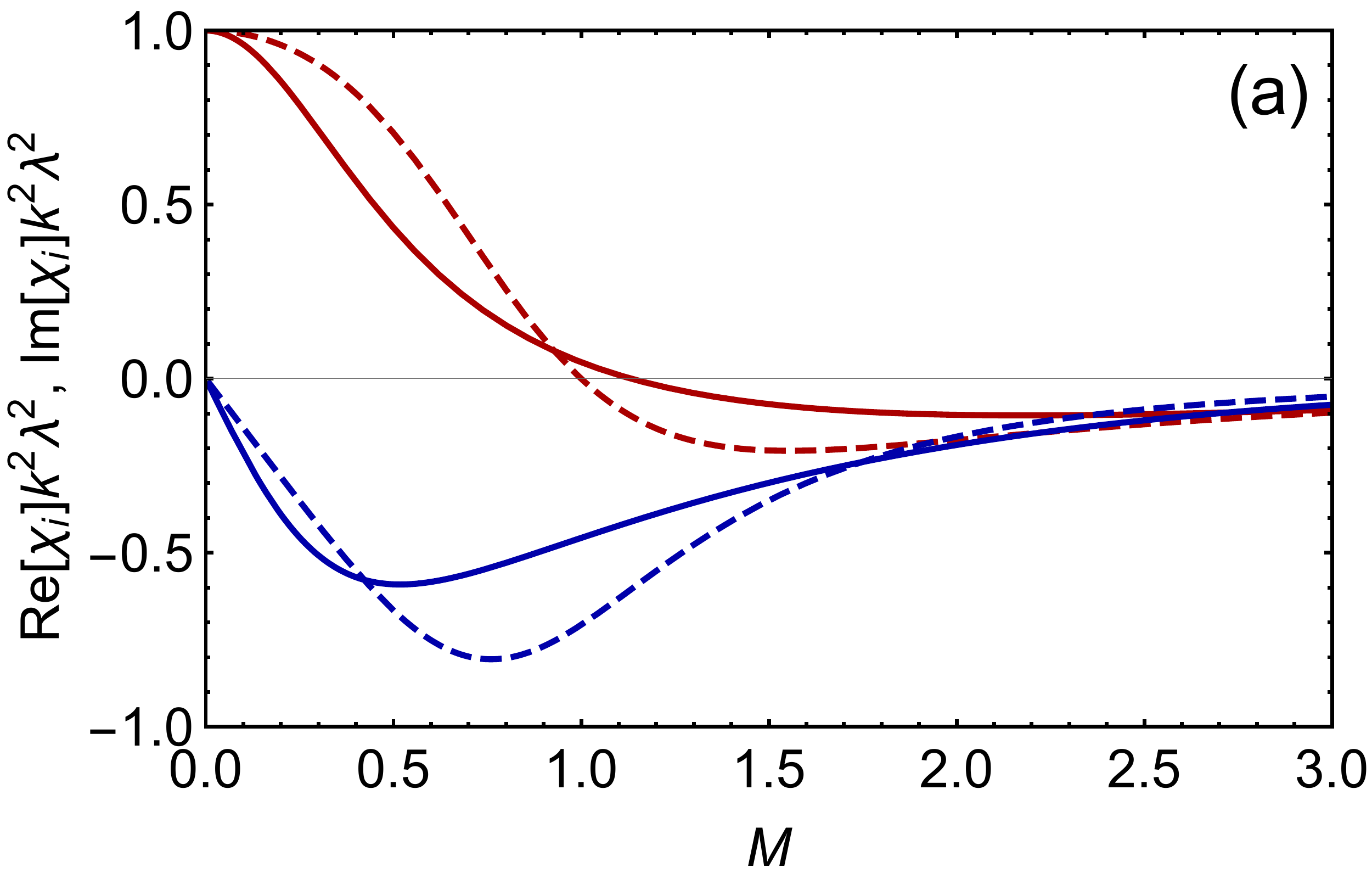} 
    \includegraphics[width=7 cm]{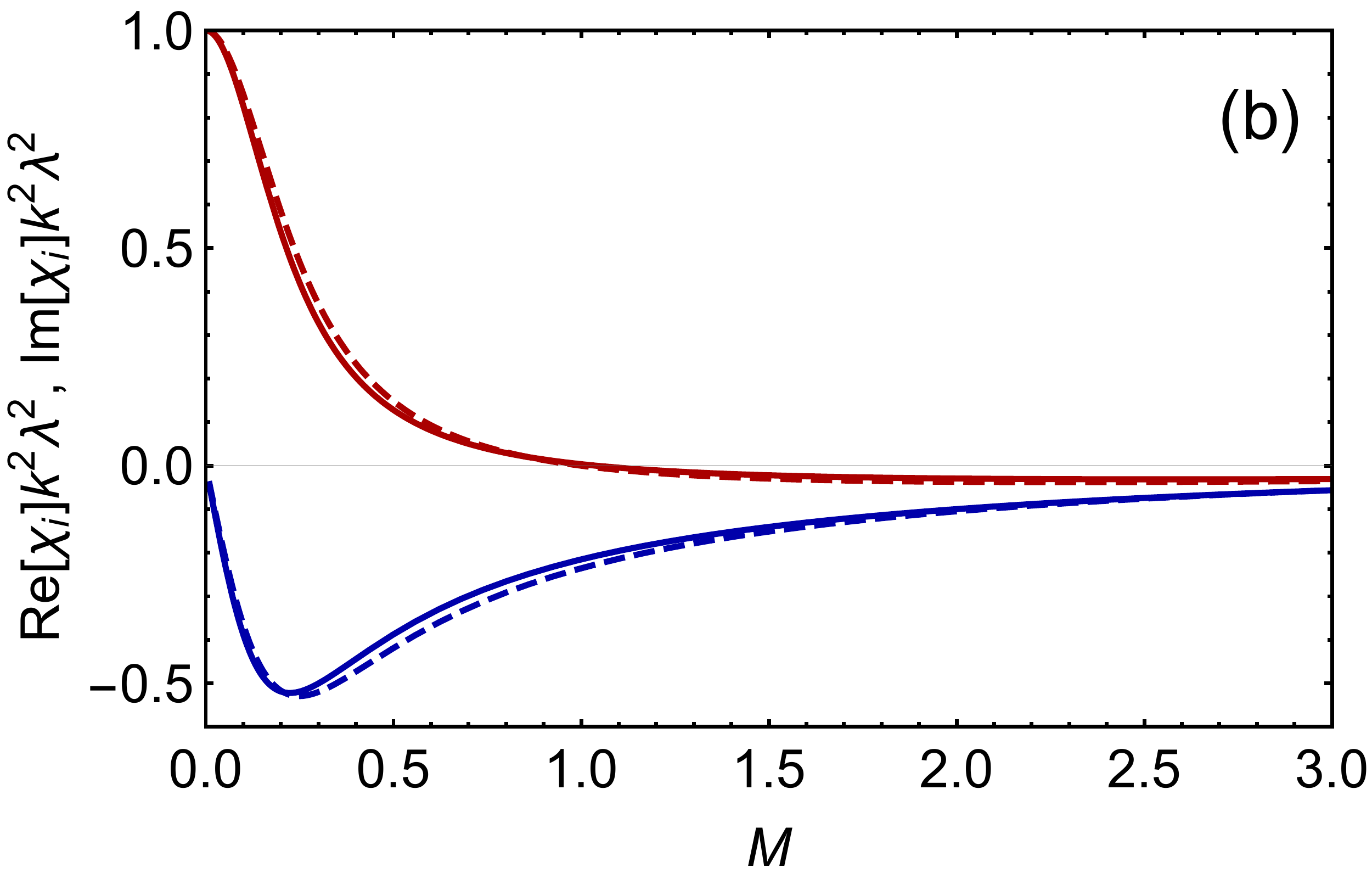} 
    \caption{Ion susceptibility versus the Mach number for two regimes of ion collisionality: $\nu_i/kv_{{\rm T}i}=\sqrt{2}$ (a) and  $\nu_i/kv_{{\rm T}i}=3\sqrt{2}$ (b). The solid curves correspond to Eq.~(\ref{ionresp1}) and the dashed curves to Eq.~(\ref{ionresp2}).}\label{Fig2}
\end{figure}

Next, we neglect dust-neutral collisions and strong coupling effects, but allow for the ion drift and ion-neutral collisions. The first useful observation is that with increasing the ion collisionality, the ion response $\chi_i$ can be well described by the conventional hydrodynamic expression~\cite{RatynskaiaPRL2004,RatynskaiaIEEE2004}
\begin{equation}\label{ionresp2}
\chi_i=-\frac{\omega_{{\rm p}i}^2}{ku(ku-i\nu_i)-k^2v_{{\rm T}i}^2}=\frac{k^{-2}\lambda_i^{-2}}{1-M^2+iM(\nu_i/kv_{{\rm T}i})}.
\end{equation}       
Figure~\ref{Fig2} illustrates this. Here we plot the ion susceptibilities ${\rm Re}[\chi_i]$ and ${\rm Im}[\chi_i]$, expressed in units $k^{-2}\lambda^{-2}$, versus the ion thermal Mach number. Figure~\ref{Fig2}(a) corresponds to the case $\nu_i/kv_{{\rm T}i}=\sqrt{2}$ and figure~\ref{Fig2}(b) to the case $\nu_i/kv_{{\rm T}i}=3\sqrt{2}$. In the second case, the difference between Eqs.~(\ref{ionresp1}) and (\ref{ionresp2}) becomes vanishingly small at all $M$. This observation can be used to simplify the description of the low frequency mode at sufficiently high ion collisionality. The conventional hydrodynamic expression becomes appropriate when the ion mean free path becomes smaller than the wavelength.   

Another related question is how the sound velocity of the low-frequency mode is affected by ion streaming. This issue has been discussed for instance in Ref.~\cite{Piel2011}, where it has been demonstrated that the phase velocity of the dust density waves can deviate greatly from the DA velocity for fast ions. This effect can be accounted for by introducing an effective screening length so that $c_{\rm s}=\omega_{\rm p}\lambda_{\rm eff}$, where $\lambda_{\rm eff}$ takes into account modifications of the plasma shielding by drifting ions. In the limit of vanishing drift we have $\lambda_{\rm eff}\simeq\lambda_i$, while in the opposite limit of very large ion drift velocity we get $\lambda_{\rm eff}\simeq \lambda_e$, since high energy ions do not contribute to screening. Various simple expressions that smoothly interpolate between these two limits have been proposed in the literature, mostly in the context of calculating the ion drag force acting on a charged particle immersed in a plasma with drifting ions~\cite{FortovPR2005,Piel2011,HutchinsonPPCF2006,KhrapakPoP2005}.    

\begin{figure}
\centering
    \includegraphics[width=7 cm]{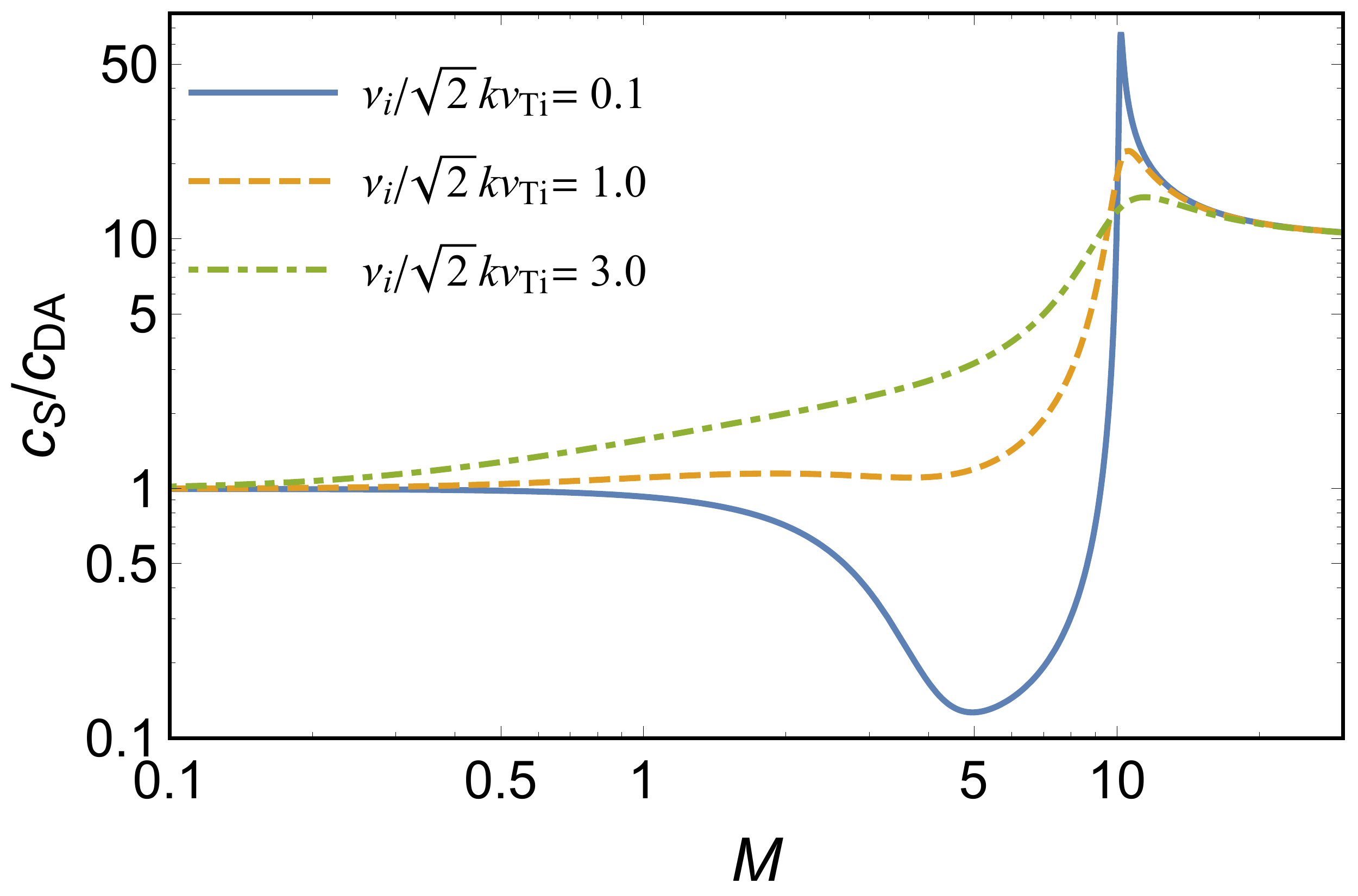} 
    \caption{Ratio of the sound velocity to the dust-acoustic velocity versus the ion thermal Mach number for three values of ion collisionality (see the legend).}\label{Fig3}
\end{figure} 

We have evaluated the low-frequency mode sound velocity in the
extended range of ion thermal Mach numbers for three different values of the ion collisionality parameter $\nu_i/kv_{{\rm T}i}$. The calculation is done for a high electron-to-ion temperature ratio, $T_e=100 T_i$, appropriate for typical gas discharge conditions. The results are shown in Fig.~\ref{Fig3}. The higher values of $M$ are not very relevant for experiments with the PK-4 laboratory, they are displayed here for completeness. Figure~\ref{Fig3} demonstrates that the actual dependence of $c_{\rm s}$ on $M$ is quite complex and non-universal. Ion drifts can either slow down the waves (at weak ion collisionality and subsonic ion drifts) or accelerate the waves (at high ion collisionality and supersonic ion drifts), with obvious implications for the effective screening length. The variations can be quite pronounced (orders of magnitude). Thus, superthermal ion drifts constitute a very important factor affecting and regulating the sound velocity of the low-frequency particle mode, along with the strongly coupling effects~\cite{KalmanPRL2000,KhrapakPRE2015_Sound,KhrapakPPCF2016},  deviations from the screened Coulomb interaction potential~\cite{RosenbergJPP2015,Fingerprints}, and space dimensionality~\cite{RosenbergJPP2015,KhrapakPoP2019}. In the limit of very high ion drift velocity, $M\gtrsim \sqrt{T_e/T_i}$, the asymptote $c_{\rm s}\simeq c_{\rm DA}\sqrt{T_e/T_i}$ will be reached, corresponding to $\lambda_{\rm eff}\simeq \lambda_{e}$ (see Fig.~\ref{Fig3}). The limiting asymptote is approached from above, independently of the ion collisionality level. This regime is not of immediate interest here, it is more relevant for sheath regions in gas discharges with supersonic ion drifts.        

\section{Instability threshold}

Now we will address the question of stability of the low-frequency particle mode, associated with the ion drift relative to the particle component. We will concentrate on subthermal ion drifts, $M\lesssim 1$. We consider three situations separately, which can be characterized as ``weakly collisional ions + weakly colisional particles'', ``strongly collisional ions + weakly collisional particles'', and `` strongly collisional ions + strongly collisional particles'', respectively. The ion collisionality with respect to particle mode propagation is characterized  by the parameter $\nu_i/kv_{{\rm T}i}$, which is roughly the ratio of the wavelength to the ion free path. The particle (dust) collisionality is characterized by the ratio of the Epstein collisional drag to the actual wave frequency, $\nu_p/ \omega$. Note that these definitions are different from those used in Sec.~\ref{parameters} above, but are more appropriate for the problem of wave propagations. In each case considered we will identify the effects which stabilize and destabilize the wave and derive rather general conditions of instability occurrence.   

\subsection{Collisionless ion regime}

This regime has been investigated by Rosenberg {\it et al}.~\cite{RosenbergPRE2014} and we consider it only briefly, for completeness.
In the first approximation, the condition ${\rm Re}[\epsilon(k,\omega)]=0$ delivers the dispersion relation of the form (\ref{sound}). The imaginary part of the ion response yields  
\begin{equation}\label{Imi}
{\rm Im}[\chi_i]\simeq -\frac{M}{k^2\lambda_i^2}\sqrt{\frac{\pi}{2}}.
\end{equation}
It destabilises the wave. The imaginary contribution from the particle component in this long-wavelength regime reads 
\begin{equation}\label{Imp}
{\rm Im}[\chi_p]\simeq \frac{\nu_p\omega_{\rm p}^2\omega}{[\omega^2-\omega_{\rm p}^2{\mathcal D}_L(k)]^2}\simeq \frac{\nu_p\omega_{\rm p}^2kc_{\rm s}}{(\omega_{\rm p}k\lambda)^4},
\end{equation}
where we have used Eq.~(\ref{sound}) along with the acoustic dispersion $\omega\simeq kc_{\rm s}$. The particles stabilize the wave through the dissipative damping associated with the particle-neutral collisions. The necessary condition for the instability is ${\rm Im}[\chi_i]>{\rm Im}[\chi_p]$. This can be rewritten as the condition on the ion thermal Mach number
\begin{equation}\label{Inst1}
M>\sqrt{\frac{2}{\pi}}\frac{\nu_p}{\omega_{\rm p}}\frac{c_{\rm s}}{c_{\rm DA}}\frac{1}{k\lambda},
\end{equation}   
where it has been assumed $\lambda_i\simeq \lambda$ for simplicity. When strong coupling effects are negligible, we have $c_{\rm s}\simeq c_{\rm DA}$ and the instability condition can be further simplified. When strong coupling effects are important, we have $c_{\rm s}< c_{\rm DA}$, and the instability condition shifts to lower $M$ (or, equivalently, the instability growth rate increases at the same $M$~\cite{RosenbergPRE2014}). The physical reason is that the magnitude of the stabilizing term ${\rm Im}[\chi_p]$, which is proportional to the actual sound velocity, decreases at strong coupling. This facilitates the instability onset and increases its growth rate. In this regime, the ion drift does not affect the sound velocity as long as $M<1$.  

In the weakly coupled regime, the condition (\ref{Inst1}) can be consistent with subthermal drift regime ($M<1$) only in the regime $\nu_p/\omega_{\rm p}< k\lambda \ll 1$.

\subsection{Collisional ion regime}

This regime corresponds to the situation where $\nu_i/kv_{{\rm T}i}\gg 1$ and, therefore, the hydrodynamic expression for the ion response (\ref{ionresp2}) can be used. The ion contribution to the imaginary part is 
\begin{equation}\label{Imi1}
{\rm Im}[\chi_i]\simeq -\frac{\nu_iMkv_{{\rm T}i}}{k^2\lambda_i^2[k^2v_{{\rm T}i}^2(1-M^2)^2+\nu_i^2M^2]}.
\end{equation}
The particle contribution is still given by Eq.~(\ref{Imp}). Consider first the case of an infinitesimally slow drift, such that $M\nu_i\lesssim kv_{{\rm T}i}$. Neglecting the difference between $\lambda$ and $\lambda_i$, and $M^2$ in comparison to unity, we can rewrite the instability condition as
\begin{equation}
\frac{\nu_i M}{v_{{\rm T}i}}\equiv \frac{eE}{T_i}\gtrsim \frac{\nu_pc_{\rm s}}{c_{\rm DA}^2},
\end{equation}          
where $E$ is the electric field. If, in addition, the particle component is in the weakly coupled (or weakly screened) regime, so that $c_{\rm s}\simeq c_{\rm DA}$ we arrive at the condition for the critical electric field, which is necessary for the instability to operate:
\begin{equation}\label{conditionE}
E\gtrsim E_{\rm cr} = \frac{T_i}{e}\frac{\nu_p}{c_{\rm DA}}.
\end{equation} 
This condition has been proposed in Ref.~\cite{FinkEPL2013} and is often quoted as the critical electric field required for
the onset of self-excited wave generation~\cite{JaiswalPoP2018,SlowWaves}. It may appear from the presented derivation that its applicability conditions are rather limited. In particular, a very slow ion drift is assumed, which is atypical in practical situations. However, we will see below that this condition can be quite useful in another parameter regime, more relevant to real experimental situations. Strong coupling effects facilitate the instability occurrence by lowering the ratio $c_{\rm s}/c_{\rm DA}$. Ion streaming effects do not affect the sound velocity at such small Mach numbers.

For a more conventional situation with $M\sim 1$ (see Fig.~\ref{Fig1}) we should take the limit $M\nu_i\gtrsim kv_{{\rm T}i}$. Neglecting again the difference between $\lambda$ and $\lambda_i$ we arrive at the instability condition of the form
\begin{equation}\label{Inst2}
k\lambda > \sqrt{\theta_i\theta_p M\frac{c_{\rm s}}{c_{\rm DA}}}.
\end{equation}       
To be consistent with the condition of long-wavelength regime $k\lambda\ll1$, the right-hand-side of Eq.~(\ref{Inst2}) should be much smaller than unity. We see again that for otherwise identical parameters strong coupling effects would facilitate the instability occurrence by reducing the factor $c_{\rm s}/c_{\rm DA}$. Note, however, that in this situation the combined effect of the ion drift and ion-neutral collisions can act in the opposite direction by enhancing the wave sound velocity (see Fig.~\ref{Fig3}).         

\subsection{Collisional particle regime}

This regime corresponds to relatively high gas pressures, so that $\nu_i/kv_{{\rm T}i}\gg 1$ and $\nu_p>\omega$. We also consider non-vanishing drifts, so that $M\nu_i\gtrsim kv_{{\rm T}i}$. In this case, the ion and particle susceptibilities are dominated by their respective imaginary parts. Equating these we readily obtain for the real frequency
\begin{equation}\label{phasevel}
\omega\simeq\omega_{\rm p}k\lambda_iM(\theta_i/\theta_p)=c_{\rm s}k.
\end{equation}  
In this case the sound velocity is completely determined by the combination of ion drift and collisional effects; strong coupling is not important in the first approximation. This mode has been discussed earlier in the literature, see for instance Ref.~\cite{KhrapakPoP12_2003}, where in addition to ion drift and collisions, the effect of particle charge variations has been retained. In connection to waves in PK-4 laboratory, this dispersion law has been invoked for the interpretation of dust density waves with discharge polarity reversal~\cite{JaiswalPoP2018,YaroshenkoPoP2019}. There are good reasons to expect that this mode is particularly relevant for the conditions of the PK-4 experiment as well as other experiments performed at moderate gas pressures.   

A stability analysis, appropriate for the regime considered, has been performed in Ref.~\cite{YaroshenkoPoP2019}. The competition between the remaining real terms in the expressions for $\epsilon(k,\omega)$ determines whether the wave is unstable or damped. 
%The electrons and ions provide a stabilizing contribution (for $M<1$), whereas the contribution from the particle component destabilizes the mode. 
The instability condition reads 
\begin{equation}\label{Inst3}
\frac{\omega_{\rm p}^2\left[\omega^2-\omega_{\rm p}^2{\mathcal D}_L(k)\right]}{\nu_p^2\omega^2}>\frac{1}{k^2\lambda_e^2}+\frac{\omega_{{\rm p}i}^2}{\nu_i^2M^2}.
\end{equation}
The necessary requirement should be satisfied 
\begin{equation}\label{Inst4}
\omega_{\rm p}^2\left[\omega^2-\omega_{\rm p}^2{\mathcal D}_L(k)\right]M^2>\omega_{{\rm p}i}^2\omega^2\frac{\nu_p^2}{\nu_i^2}.
\end{equation}  
If this condition is satisfied, then equation (\ref{Inst3}) can be used to estimate the critical wavenumber above which the instability can operate, similarly to how this was done for a weakly coupled complex plasma in Refs.~\cite{KhrapakPoP12_2003,YaroshenkoPoP2019}. 

In the weakly coupled regime, the requirement (\ref{Inst4}) reduces to
\begin{equation}\label{inst5}
{\mathcal H}=M(\theta_i/\theta_p) >1.
\end{equation}   
We observe that ${\mathcal H}$ is a very important parameter, which in the considered regime determines the sound velocity of the low-frequency particle mode as well as its instability condition. The condition (\ref{inst5}) can be easily met in real experiments at sufficiently low pressures (see Fig~\ref{Fig1} for an example). It implies that the wave phase velocity exceeds the conventional DA velocity. Interestingly enough, rewritten in the appropriate form, this condition coincides with that given in Eq.~(\ref{conditionE}), another demonstration of its general importance.   Effects of strong coupling further facilitate the instability occurrence, because ${\mathcal D}_L(k)<0$ and thus the term responsible for wave destabilization increases at strong coupling. 

\section{How to estimate ${\mathcal D}_L(k)$?}\label{D_L}

The rather general expressions derived in the preceding sections have demonstrated that the effect of strong coupling can affect (via ${\mathcal D}_L(k)$) both the sound velocity of the low-frequency particle mode and its growth or dissipation rate. Simple practical tools to evaluate ${\mathcal D}_L(k)$ would be helpful. There are no principle difficulties in obtaining accurate RDFs (via various integral equation schemes or via the direct molecular dynamics (MD) or Monte Carlo (MC) simulations~\cite{OttPoP2014,ToliasPRE2014,ToliasPoP2019}), and then performing numerical integration in Eq.~(\ref{QLCA}). However, there are much simpler approaches, which deliver essentially the same accuracy in the considered long-wavelength limit. 

One particularly simple approach is to use a simplest one-step approximation for the RDF~\cite{KhrapakPoP2016}. In this approximation $g(r)=\theta(r-R)$, where $\theta(x)$ is the Heaviside step function, and $R$ is the correlational hole radius, which accounts for strong interparticle repulsion at short separations. The latter is a free adjusting parameter, which can be determined from the condition that the internal energy (or pressure) are correctly evaluated from this one-step approximation. This simple model is particularly appropriate for soft repulsive potentials, where the cumulative contribution from long distances (where $g(r)$ oscillates around unity) dominates~\cite{KhrapakPoP2016Log,KhrapakPRE2018,KhrapakJCP2018,KhrapakJCP2019} and for the long-wavelength regime considered here.

\begin{figure}
\centering
    \includegraphics[width=7 cm]{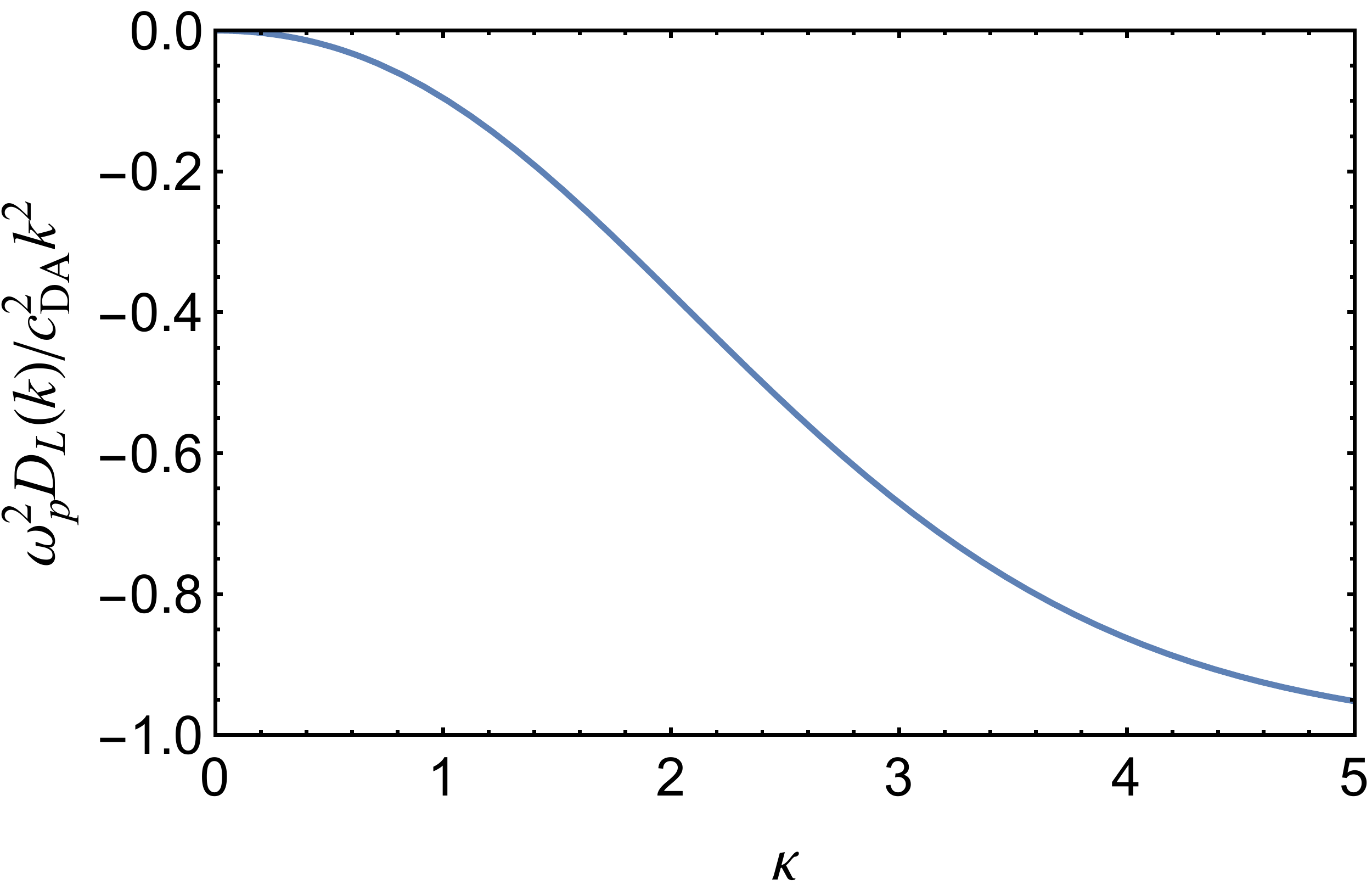} 
    \caption{Reduced longitudinal projection of a dynamical matrix, $\omega_{\rm p}^2{\mathcal D}_L(k)/c_{\rm DA}^2k^2$, in the long-wavelength regime versus the screening parameter $\kappa$.}\label{Fig4}
\end{figure} 

With this simple model RDF, the integral in Eq.~(\ref{QLCA}) is easily evaluated analytically~\cite{KhrapakPoP2016,KhrapakIEEE2018}. In the long-wavelength limit the we can obtain:
\begin{equation}
\omega_{\rm p}^2{\mathcal D}(k)=c_{\rm DA}^2k^2\left[f(\kappa R)-1\right],
\end{equation}  
where $\kappa$ is the screening parameter, $\kappa=(4\pi n_p/3)^{-1/3}/\lambda$, and the function $f(x)$ is defined via~\cite{KhrapakAIPAdv2017}
\begin{displaymath}
f(x)=e^{-x}\left(1+x+\frac{13}{30}x^2+\frac{1}{10}x^3\right).
\end{displaymath}
A good empirical approximation for the correlational hole radius in the strongly coupled regime is provided by 
\begin{equation}
R(\kappa)\simeq 1.0955+\tfrac{1}{10}\kappa-\tfrac{1}{180}\kappa^2.
\end{equation}
The dependence $f(\kappa R)-1$ on $\kappa$ is plotted in Fig.~\ref{Fig4}. We observe that the magnitude of strong coupling effects is relatively small at weak screening ($\kappa\lesssim 1$), but becomes quite important at $\kappa>1$.  

Similar results could be obtained with a two-step approximation for the RDF, discussed recently~\cite{Fairushin}. Alternatively, since in the long-wavelength regime we can write $\omega_{\rm p}^2{\mathcal D}_L(k)\simeq (c_{\rm s}^2-c_{\rm DA}^2)k^2$, we can use a thermodynamic consideration to calculate $c_{\rm s}$~\cite{KhrapakPRE2015_Sound,KhrapakPPCF2016}, based on the knowledge of an accurate equation of state of Yukawa fluids~\cite{KhrapakPRE2015,KhrapakJCP2015}.
         
\section{Numerical example}  

\begin{figure}
\centering
    \includegraphics[width=7 cm]{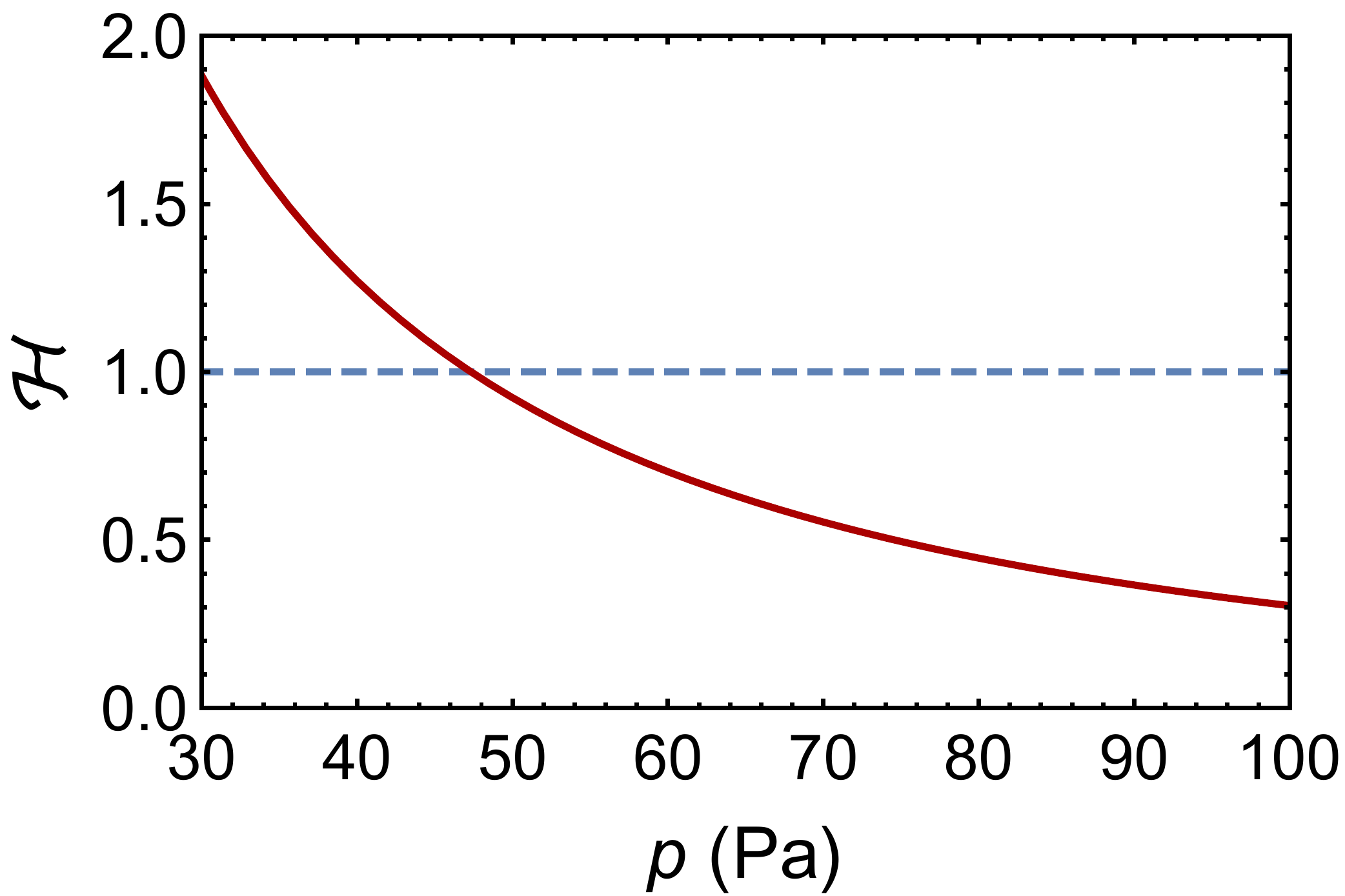} 
    \caption{The dimensionless parameter ${\mathcal H}=M(\theta_i/\theta_p)$ versus the neutral gas pressure for the parameters summarized in Section~\ref{parameters}.}\label{Fig5}
\end{figure}

Let us use the parameter regime of Section~\ref{parameters} to provide an illustration of the obtained results. For simplicity we fix the particle density and treat the neutral gas pressure as the varying parameter. Figure~\ref{Fig5} plots the parameter ${\mathcal H}$ versus the neutral gas pressure. Condition (\ref{inst5}) is satisfied for $p\lesssim 50$ Pa, which can be used as a first approximation for the instability threshold. 

\begin{figure*}[htb]
\centering
    \includegraphics[width=17 cm]{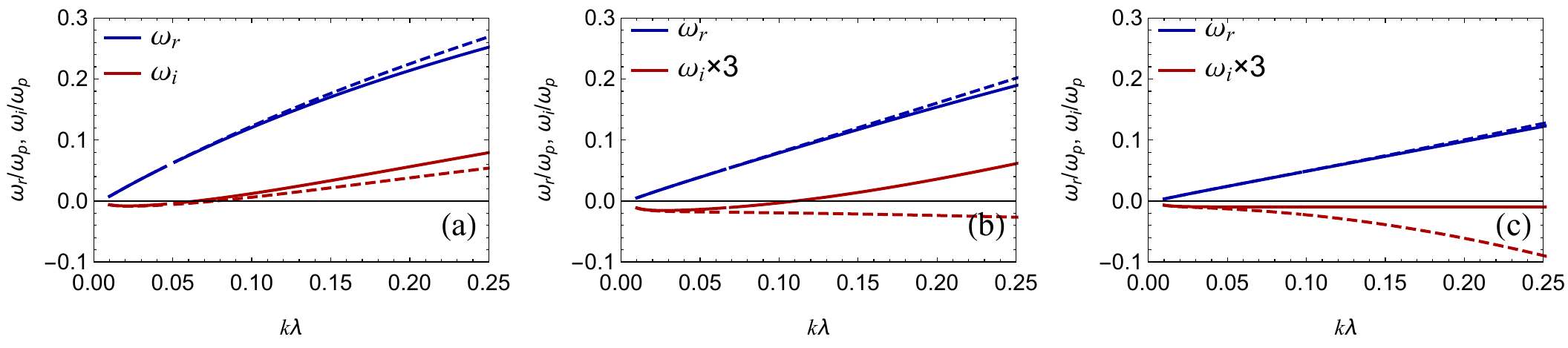} 
    \caption{Dispersion relations of the low-frequency particle wave propagating along the direction of the ion flow for three different neutral gas pressures: $p=40$ Pa (a),  $p=55$ Pa (b), and  $p=75$ Pa (c). Other plasma parameters are evaluated using the approximations from Section~\ref{parameters}. Upper (blue) curves correspond to the real component of the wave frequency, $\omega_r$. Lower (red) curves show  the imaginary component of the frequency, $\omega_i$. Solid curves correspond to the calculation with strong coupling effects included, dashed curves show the calculatiion without strong coupling effects. Note that in (b) and (c) the imaginary component is multiplied by a factor 3 for clarity. The range of reduced wavenumbers $k\lambda$ shown corresponds to the long-wavelength limit, considered in this work. For the discussion see the text.}\label{Fig6}
\end{figure*}

Figure~\ref{Fig6} presents three examples of the long-wavelength dispersion relations of the low-frequency mode propagating along the ion flow for the plasma parameters discussed in Section~\ref{parameters}. These are different by the neutral gas pressure, which increases from $p=40$ Pa in (a), through $p=55$ Pa in (b), to $p=75$ Pa in (c). The dispersion relations have been calculated numerically by equating (\ref{permit}) to zero with the electron response from Eq.~(\ref{electronresp}), the ion response from the hydrodynamic expression (\ref{ionresp2}), and the particle response from Eq.~(\ref{partresp}). The electron-to-ion temperature ratio is fixed at $T_e/T_i=100$. Solid curves correspond to retained strong coupling effects, dashed curves to the calculation without strong coupling. We observe that for the case investigated the real frequency is not very sensitive to the strong coupling effects. The wave phase velocity decreases with pressure, in agreement with Eq~(\ref{phasevel}) and Fig~\ref{Fig5}. The imaginary part of the frequency is much more sensitive to the strong coupling effects. At $p=40$ Pa the wave is unstable, independently of whether strong coupling effects are included or not. Strong coupling effects enhance the instability growth rate. At $p=55$ Pa the mode is only unstable when strong coupling effects are accounted for. At $p=75$ Pa the mode becomes damped, independently of whether strong coupling effects are included or not. Thus, Eq.~(\ref{inst5}) can be considered as an appropriate first guess for the instability threshold. At the same time, strong coupling effects can shift the threshold to considerably higher pressures.         

\section{Conclusion}     

We have investigated the properties of a low-frequency particle mode and its instability in a collisional complex plasma with drifting ions, taking into account the effects of strong coupling in the particle component. We have considered sufficiently long wavelengths, for which acoustic dispersion takes place. First, we have studied the effect of ion drift on the phase (sound) velocity of the mode. It has been demonstrated that for a superthermal ion drift, the ion flow can either slow down or accelerate the wave (compared to the conventional isotropic DA wave), depending on the ion collisionality. Then, considering subthermal ion drifts, a general condition of the mode instability has been derived. This extends the analysis of Ref.~\cite{RosenbergPRE2014}, which is limited to the weakly collisional ion regime, to the regime where the ion mean free path with respect to collisions with neutrals is shorter than the mode wavelength. We observe that quite generally strong coupling effects promote the instability by simplifying its occurrence and enhancing its growth rate. We give a recipe on how to evaluate strong coupling effects numerically. A numerical example is presented to clearly illustrate the obtained results.   
The wide parameter regime considered in this study is representative to the PK-4 experiment, currently operational on board the International Space Station, as well as many other laboratory experiments with complex plasmas in weak external electric fields. The obtained results should be useful in analysing and interpreting observations in these experiments.   

\acknowledgments

We thank Mierk Schwabe for reading the manuscript. Work on dust density waves in the PK - 4 laboratory was supported by the Russian Science Foundation, Grant No. 20-12-00365.

\bibliographystyle{aipnum4-1}
\bibliography{Instability}

\end{document}